\documentclass[conference,11pt]{IEEEtran}

\usepackage{balance}
\usepackage{amsmath}
\usepackage{graphicx}

\newtheorem{theorem}{Theorem}[section]
\newtheorem{ex}{Example}[section]

\newcommand{\bm}[1]{\mbox{\boldmath{$#1$}}}

\begin{document}

\title{Constructing CSS Codes with LDPC Codes for the BB84 Quantum Key Distribution Protocol}

\author{\IEEEauthorblockN{\Large Maki~Ohata\IEEEauthorrefmark{1} and Kanta Matsuura\IEEEauthorrefmark{1}}
\IEEEauthorblockA{\IEEEauthorrefmark{1}Department of Information and Communication Engineering\\
Graduate School of Information Science and Technology\\
The University of Tokyo\\
4-6-1 Komaba, Meguro-ku, Tokyo, 152-8505 Japan.\\
Email: \{ohata, kanta\}@iis.u-tokyo.ac.jp}
}

\maketitle

\begin{abstract}
%\boldmath
In this paper, we propose how to simply construct a pair of linear codes for the BB84 quantum key distribution protocol. 
This protocol allows unconditional security in the presence of an eavesdropper, 
and the pair of linear codes is used for error correction and privacy amplification.
Since their high decoding performance implies low eavesdropper's mutual information, 
good design of the two codes is required. 
The proposed method admits using arbitrary low-density parity-check (LDPC) codes. 
Therefore, it has low complexity and high performance for hardware implementation.  
Simulation results show that the pair of codes performs well against practical and various noise levels.
\end{abstract}


\IEEEpeerreviewmaketitle

\section{Introduction}
Public-key cryptosystems, based on some computational assumption, such as RSA \cite{RSA78} 
are becoming weaker due to the progress of cryptanalysis and computing power, although long-term security has been demanded for diverse situations. 
In addition, if realistic quantum computers appeared, most current public-key cryptosystems would be broken \cite{S97}. 

On the other hand, a one-time pad offers information theoretic security.
Due to random outputs that bear no statistical relationship to plaintexts, 
an adversary cannot obtain information about plaintexts from ciphertexts. 
A big problem with the one-time pad is that it requires to distribute truly random keys of the same length as plaintexts, at a rate of one per plaintext. 

One solution for the key distribution problem was suggested by Bennett and Brassard in 1984 \cite{BB84}. 
In the presence of an eavesdropper (Eve) with unlimited computational power, 
the BB84 quantum key distribution protocol allows unconditional security under the sole assumption that the laws of physics are correct \cite{M01,SP00}. 

In the BB84 protocol, a pair of classical binary linear codes, known as Calderbank-Shor-Steane (CSS) codes \cite{CSS96}, 
is used for error correction and privacy amplification over a classical channel when a quantum channel is noisy.  
With the use of their decoding error probability, Eve's mutual information can be bounded. 

Quantum codes, applicable as CSS codes, have been studied previously by employing classical codes such as Hamming, BCH, and Reed-Solomon codes \cite{CRSS98,GGB99}. 
These classic codes have plainly constructible dual ones for composing CSS codes, 
but their decoding performance is insufficient to be decodable against practical noise levels over the quantum channel. 
Consequently, it is necessary to design and evaluate the better pair of codes. 
In theoretical, after fixing one linear code, randomly choosing the other one is recognized as a good construction approach \cite{WMU04}. 
However, a non-random and practical method for designing the proper pair of codes remains an open problem. 

Low-Density Parity-Check (LDPC) codes \cite{G63} are a class of error-correcting linear block codes 
admitting representations in terms of sparse bipartite graphs, known as Tanner graphs \cite{T81}, with variable nodes and check nodes. 
LDPC codes provide near capacity performance on memoryless channels by the sum-product decoding algorithm. 
The purpose of this paper, therefore, is to construct the good pair of codes using LDPC codes. 
 
MacKay {\it et al.} have shown how to create the two codes by means of dual-containing LDPC codes \cite{MMM04}.
In the dual-containing LDPC codes, every pair of rows in their parity-check matrices must have an even overlap, and every row must have even weight.
Because of these properties, the parity-check matrices have many cycles of length four in Tanner graphs. 
Since short cycles, particularly length four, in Tanner graphs have a bad influence on decoding performance by the sum-product algorithm, 
the technique presented in \cite{MMM04} cannot apply optimal LDPC codes. It is suitable for quantum error-correcting codes but not for the BB84 protocol.

In this paper, we propose the simple and practical method for constructing the pair of codes admitting the use of arbitrary LDPC codes. 
For this reason, our proposal does not need high complexity for hardware implementation, and 
the pair of codes designed by the proposed method has the same decoding performance as chosen, namely, optimal LDPC codes. 

This paper is organized as follows. 
In Section \ref{BB84} we introduce the BB84 quantum key distribution protocol. 
In Section \ref{LDPC} we describe efficiently encodable irregular LDPC codes. 
In Section \ref{D} we propose how to construct the two codes using LDPC codes. 
In Section \ref{sim} we discuss the performance evaluation and show several simulation results for the proposed method. 
In Section \ref{con} we summarize the results.

\section{BB84 Quantum Key Distribution Protocol}
\label{BB84}
\subsection{CSS Codes}
CSS codes invented by Calderbank, Shor, and Steane \cite{CSS96} are a class of quantum error-correcting codes, 
and they are derived from classical linear codes by using the concept of dual codes. 
 
We first describe CSS codes. Let $C_1$ and $C_2$ be an $[n,k_1]$ and $[n,k_2]$ classical linear code such that $C_1\supset C_2$.  
With the use of $C_1$ and $C_2$, basis vectors for the CSS code, denoted as an $[n,k_1-k_2]$ quantum code, subspace can be represented as 
$$\bm{v} \longrightarrow \frac{1}{\sqrt{|C_2|}} \sum_{\bm{w} \in C_2} |\bm{v}+\bm{w}\rangle,$$
where $\bm{v} \in C_1$. 
If the code $C_2$ includes a codeword $\bm{v}_1-\bm{v}_2$, then the codewords corresponding to $\bm{v}_1$ and $\bm{v}_2$ are the same. 
Thus, these codewords are equivalent to cosets of $C_2$ in $C_1$. 
The code $C_1$ is used for bit-flip error correction, and the code $C_2^\perp$ is done for phase-flip one after the application of an Hadamard transform. 

\subsection{Procedure of the BB84 Protocol}
In this subsection, for simplicity, we omit the procedure of the BB84 protocol over the quantum channel and 
introduce that over the classical channel. 

In the BB84 protocol, a sender (Alice) and a receiver (Bob) are participants. 
After communicating over the quantum channel and estimating a noise rate, Alice has an $n$ bit string $\bm{x}$, and Bob has an $n$ bit string $\bm{x} + \bm{e}$.   
The procedure over the classical channel is as follows: 
\begin{enumerate}
\item Alice chooses a random $n$ bit codeword \bm{u} $\in C_1$.
\item Alice announces $\bm{x} + \bm{u}$ to Bob.  
\item Bob subtracts $\bm{x} + \bm{u}$ from $\bm{x} + \bm{e}$ and obtains $\bm{u}'$ by error-correcting the result, $\bm{u} + \bm{e}$, to the codeword in $C_1$.
\item Alice obtains the coset of $\bm{u} + C_2$ as keys, and Bob obtains the coset of $\bm{u}' + C_2$ as keys.
\end{enumerate}
The code pair of $C_1$ and $C_2$ is CSS codes. In the BB84 protocol, it is used for error correction and privacy amplification.  
For correcting errors, it requires that the code $C_1$ can be decodable by some practicable decoding algorithm. 
Even though Alice's codeword $\bm{u}$ is not equal to Bob's codeword $\bm{u}'$, 
provided that $\bm{u}'-\bm{u} \in C_2$, Alice and Bob can share the same keys. 
Thus, for sharing the keys with high probability, the performance of the coset $C_1/C_2$ is important rather than that of the code $C_1$. 

\subsection{Security Evaluation of the BB84 Protocol}
An entanglement purification protocol can bound Eve's mutual information on the shared keys \cite{LC99}. 
If Alice and Bob share a $k$-EPR-pair state with high fidelity $F$, where $F>1-\delta$, then Eve's mutual information can be bounded by 
$$I_{Eve}<-(1-\delta ) \log_2(1-\delta )-\delta \log_2\frac{\delta }{2^{2k}-1}.$$
This inequality shows that high fidelity implies low entropy. 
By reduction from the entanglement purification protocol to the BB84 protocol via CSS codes \cite{SP00}, the parameter $\delta$ corresponds to the decoding error probability of 
the worse of the two codes $C_1(C_1/C_2)$ and $C_2^{\perp}(C_2^\perp /C_1^\perp)$. 
Hence, high decoding performance implies low Eve's mutual information, 
and the decoding error probability of the two codes must be small. 
In this paper, we analyze the {\it block} error probability.

\section{LDPC Codes}
\label{LDPC}
Originally invented by Gallager in the early 1960's \cite{G63}, LDPC codes have greatly developed as one of the most promising error-correcting codes in the last few years. 
They have been recently adopted as standard error-control coding techniques in several communication systems. 

\begin{figure}[ht]
\begin{center}
\includegraphics[scale=0.28]{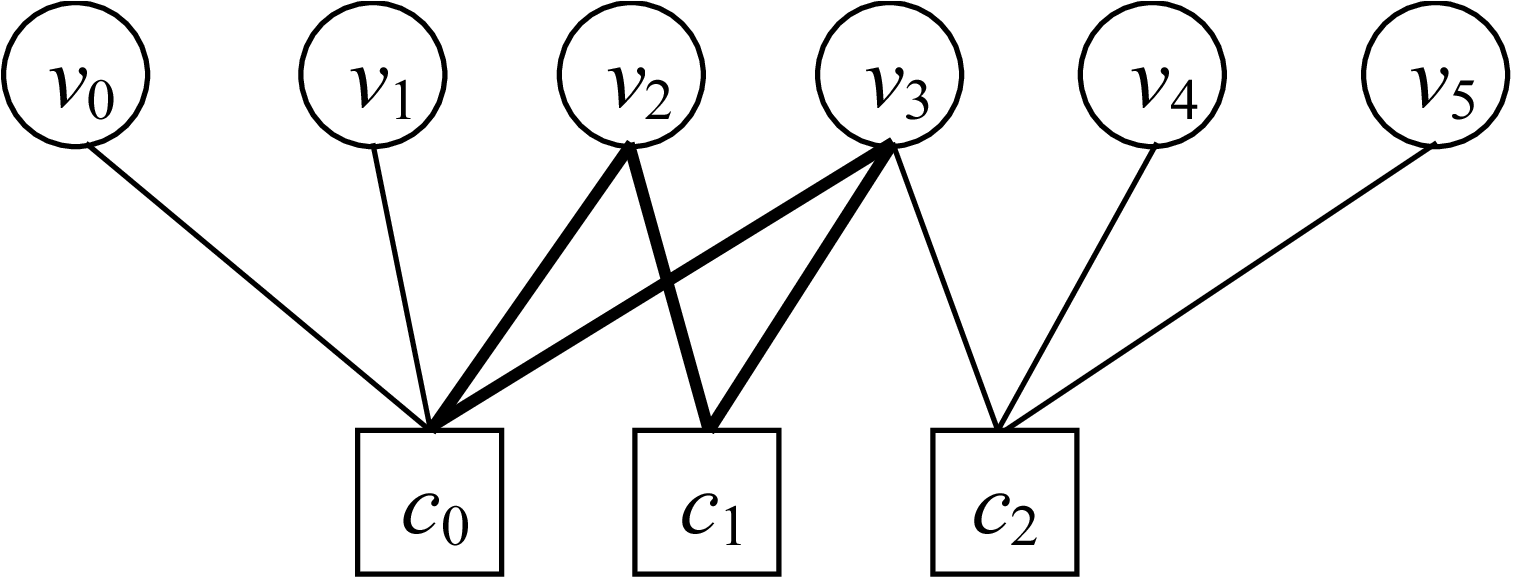}\\
 \caption{A Tanner graph}
\label{tan}
\end{center}
\end{figure}

Every binary linear code of length $n$ and dimension $k$ can be represented as Tanner graphs with variable nodes $v_i,0\leq i\leq n-1$ and check nodes $c_j,0\leq j\leq k-1$. 
The following is an example for this representation:
\begin{ex}
Assume that the parity-check matrix $H$ of LDPC codes of length $n=6$ and dimension $k=3$ is given by
\begin{equation*}
H= 
\begin{bmatrix}
\hspace{2mm}1&1&1&1&0&0\hspace{2mm} \\
\hspace{2mm}0&0&1&1&0&0\hspace{2mm} \\
\hspace{2mm}0&0&0&1&1&1\hspace{2mm}
\end{bmatrix}.
\end{equation*}
The Tanner graph representing this LDPC code is shown in Fig. \ref{tan}D
A cycle of length four can be seen in Fig. \ref{tan} that the set \{$v_2,v_3,c_0,c_1$\} marked by bold lines. 
Short cycles in Tanner graphs degrade the decoding performance by the sum-product algorithm. 

\end{ex}

LDPC codes are classified into regular LDPC codes and irregular ones. 
The decoding performance of irregular LDPC codes is superior to that of regular ones, 
but irregular LDPC codes usually need high complexity for their encoding processes. 

Irregular LDPC codes presented in \cite{FOS05} have algebraic structure and an efficient encoding algorithm. 
Besides, they perform well as compared to randomly constructed irregular LDPC codes.
In this paper, therefore, we choose them as the code $C_1$. 

\subsection{Efficiently Encodable Irregular LDPC Codes}
In this subsection, we introduce how to construct irregular LDPC codes proposed by Fujita {\it et al.} \cite{FOS05}. 
Let $p$ be an odd prime number such that $2\leq j\leq k\leq p-1$, and let $H$ be a parity-check matrix defined by an $M(:=pj) \times N(:=pj+pk)$ matrix $[H^{(p)}~|~H^{(d)}]$, 
where the $M \times M$ submatrix $H^{(p)}$ and the $M \times (N-M)$ submatrix $H^{(d)}$ are defined in block form as follows:
\begin{equation*}
H^{(p)}:= 
\begin{bmatrix}
\hspace{.9mm}   T\hspace{-1mm} & I\hspace{-1mm} & O\hspace{-1mm} & \cdots \hspace{-1mm} & O\hspace{-1mm} & O\hspace{.9mm}  \\
\hspace{.9mm}   O\hspace{-1mm} & I\hspace{-1mm} & I\hspace{-1mm} & \cdots & O\hspace{-1mm} & O\hspace{.9mm}   \\
\hspace{.9mm}   O\hspace{-1mm} & O\hspace{-1mm} & I\hspace{-1mm} & \cdots & O\hspace{-1mm} & O\hspace{.9mm}    \\
\hspace{.9mm}   \vdots  \hspace{-1mm}& \vdots \hspace{-1mm}& \vdots \hspace{-1mm}&  \ddots \hspace{-1mm}&  \vdots \hspace{-1mm} & \vdots \hspace{.9mm}    \\
\hspace{.9mm}   O\hspace{-1mm} & O\hspace{-1mm} & O\hspace{-1mm} & \cdots & I\hspace{-1mm} & I\hspace{.9mm}    \\
\hspace{.9mm}   O \hspace{-1mm}& O\hspace{-1mm} & O\hspace{-1mm} & \cdots & O\hspace{-1mm} & I\hspace{.9mm}  
\end{bmatrix},
\end{equation*}
\begin{equation*}
H^{(d)} :=
\begin{bmatrix}
\hspace{1.2mm} I &  I & \cdots  & I \hspace{.5mm} \\
\hspace{1.2mm}  P & P^2   & \cdots &  P^{k} \hspace{.5mm} \\
\hspace{1.2mm}   P^2 & P^4 & \cdots &  P^{2k} \hspace{.5mm} \\
\hspace{1.2mm} \vdots & \vdots &  \vdots &  \vdots \hspace{.5mm} \\
\hspace{1.2mm}  P^{j-1} & P^{2(j-1)}  & \cdots & P^{k(j-1)} \hspace{.5mm}
\end{bmatrix}, 
\end{equation*}
where $I$ is the $p \times p$ identity matrix, $O$ is the $p \times p$ matrix of zeros;  
the $p \times p$ matrices $P$ and $T$ are defined by the following matrices:
\begin{equation*}
P :=
\begin{bmatrix}
\hspace{.9mm} 0\hspace{-1mm}& 0 \hspace{-1mm}& \cdots \hspace{-1mm}& 0 \hspace{-1mm}& 1\hspace{.9mm} \\
\hspace{.9mm} 1\hspace{-1mm}& 0 \hspace{-1mm}& \cdots \hspace{-1mm}& 0 \hspace{-1mm}& 0\hspace{.9mm} \\
\hspace{.9mm} 0 \hspace{-1mm}& 1 \hspace{-1mm}&\cdots \hspace{-1mm}& 0 \hspace{-1mm}& 0\hspace{.9mm} \\
\hspace{.9mm} \vdots \hspace{-1mm} &  \vdots  \hspace{-1mm}& \ddots \hspace{-1mm}&  \vdots \hspace{-1mm}&  \vdots \hspace{.9mm} \\
\hspace{.9mm} 0 \hspace{-1mm}& 0 \hspace{-1mm}& \cdots \hspace{-1mm}& 1 \hspace{-1mm}& 0 \hspace{.9mm} 
\end{bmatrix},
\end{equation*}
\begin{equation*}
T:= 
\begin{bmatrix}
\hspace{.9mm}   1\hspace{-1mm} & 1\hspace{-1mm} & 0\hspace{-1mm} & \cdots \hspace{-1mm} & 0\hspace{-1mm} & 0\hspace{.9mm}  \\
\hspace{.9mm}   0\hspace{-1mm} & 1\hspace{-1mm} & 1\hspace{-1mm} & \cdots & 0\hspace{-1mm} & 0\hspace{.9mm}  \\
\hspace{.9mm}   0\hspace{-1mm} & 0\hspace{-1mm} & 1\hspace{-1mm} & \cdots & 0\hspace{-1mm} & 0\hspace{.9mm} \\
\hspace{.9mm}   \vdots  \hspace{-1mm}& \vdots \hspace{-1mm}& \vdots \hspace{-1mm}&  \ddots \hspace{-1mm}&  \vdots \hspace{-1mm} & \vdots \hspace{.9mm} \\
\hspace{.9mm}   0\hspace{-1mm} & 0\hspace{-1mm} & 0\hspace{-1mm} & \cdots & 1\hspace{-1mm} & 1\hspace{.9mm}   \\
\hspace{.9mm}   0 \hspace{-1mm}& 0\hspace{-1mm} & 0\hspace{-1mm} & \cdots & 0\hspace{-1mm} & 1\hspace{.9mm}  
\end{bmatrix}.
\end{equation*} 
In addition, optimal irregular LDPC codes are constructed by appropriately changing several block component of $H^{(d)}$ into the $p \times p$ matrix $O$ (called a masking method \cite{CDXLA04}). 

\begin{figure}[!t]
\begin{center}
\includegraphics[scale=0.685]{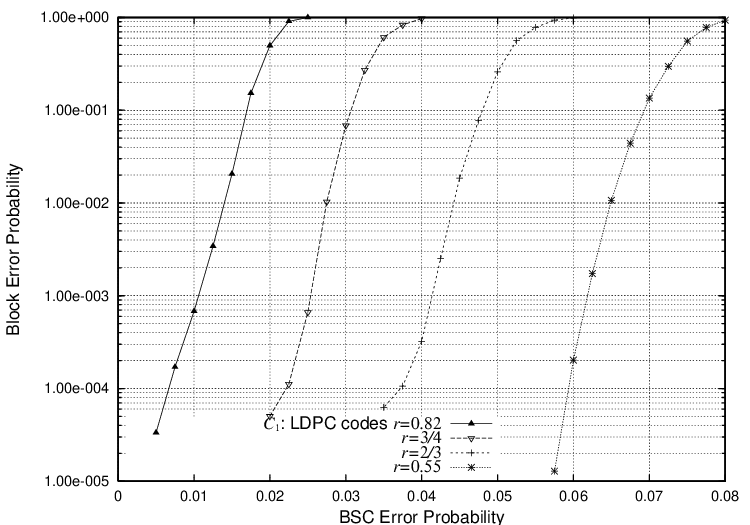}
 \caption{Decoding performance of irregular LDPC codes of length almost 5000 and several rates on the BSC.}
\label{air}
\end{center}
\end{figure}

Fig. \ref{air} depicts the performance result of almost length 5000 and rate 0.82, 3/4, 2/3, and 0.55 by sum-product decoding with up to 100 iterations on a binary symmetric channel (BSC). 
In all cases, we employed $p=73$ as the prime number, and 
their LDPC codes were designed based on masking matrices given in App. \ref{AA}. 
Despite the negligible area of probability for error correction, the LDPC codes of rate 3/4 and 2/3 have an error floor starting at the block error probability of $10^{-4}$ 
as compared to those of rate 0.82 and 0.55 due to column wight three in their party-check matrices. 
However, due to little information about the detailed methods for constructing optimal masking matrices, 
we note that their elaborate design can further improve their performance.

\section{Design of CSS Codes}
\label{D}
In this section, we propose how to construct the pair of codes using LDPC codes. the method is executed by the following procedure: 
\begin{enumerate}
\item Choose arbitrary LDPC codes defined by an $M \times N$ parity-check matrix $H_1$ for the code $C_1$.
\item Separate an $M\times (N-M)$ matrix $H_1'$ from $H_1$ in ascending order of column weights, 
encode bits of length $N-M$, where these bits are the row vectors of $H_1'$, to codewords in $C_1$, and 
generate the $M \times N$ parity-check matrix $H_2$ by the set of the codewords of $C_1$. 
\item Defined the code $C_2^{\perp}$ by $H_2$.
\end{enumerate}
The parity-check matrix $H_2$ is composed of codewords in $C_1$, we have $H_1H_2^T=O$, hence the two codes satisfy the CSS code condition $C_1\supset C_2$. 

This method can apply arbitrary LDPC codes and simply construct the two codes by only choosing the code $C_1$. 
By using pseudo-random property of LDPC codes, it pseudo-randomly creates the code $C_2$. 
In short, the proposed method is a practical approach of the random construction presented in \cite{WMU04}. 

The parity-check matrix $H_2$ has a low-density submatrix and a high-density submatrix. 
As for theoretical decoding performance, 
since linear codes defined by high-density parity-check matrices have better performance than those by low-density ones \cite{SU03}, 
it can be anticipated that the code $C_2^{\perp}$ has a superior decoding property.     

Suppose that the rate of $C_1$ is $r$, the rate of $C_2^{\perp}$ is also $r$, 
so the rate as CSS codes designed by this method is $2r-1$.

\section{Simulation Results}
\label{sim}
In this section, we show the simulation results for the pair of linear codes constructed by the propose method on the BSC. 
We can consider errors between Alice and Bob as ones over the BSC. 
First of all, we discuss how to evaluate the decoding performance of the code $C_2^{\perp}(C_2^\perp /C_1^\perp)$. 

\subsection{Decoding Performance Analysis of the Code $C_2^{\perp}$}
The code $C_1$ is LDPC codes; therefore, its performance can be simply evaluated by sum-product decoding.  
Since the code $C_2^{\perp}$ does not directly correct errors in the BB84 protocol, 
it is sufficient to analyze the performance by some robust and specifically evaluable decoding algorithm. 

While maximum likelihood decoding (MLD) is a powerful algorithm, we cannot evaluate the performance of $C_2^{\perp}(C_2^\perp /C_1^\perp)$ due to its decoding complexity. 
Moreover, we cannot analyze the performance by the general sum-product decoding algorithm as well because of a high-density part of $H_2$. 
Owing to the difficulty of the decoding performance evaluation, 
both codes tend to be constructed as mediocre LDPC codes and evaluated by sum-product decoding \cite{HI06,LG05,MMM04}. 

As a robust decoding algorithm having close performance to MLD, Fossorier {\it et al.} have proposed to combine sum-product decoding with ordered statistic decoding (OSD) \cite{FL95}.   
In the present paper, instead of general sum-product decoding, 
we applied bit serial sum-product decoding \cite{YKYK04}, which is the basically same algorithm as shuffled belief propagation decoding \cite{ZF02}, 
and we remove cycles of length four \cite{KSS04,YCF02} for solving short cycles in Tanner graphs prior to decoding.  
Bit serial sum-product decoding \cite{YKYK04} is the improved algorithm, operating the propagation of likelihood information in Tanner graphs bit-by-bit, of general sum-product decoding, and 
the algorithm in \cite{KSS04,YCF02} can remove cycles of length four by transforming Tanner graphs under the code-equivalent condition. 

\begin{figure}[!t]
\begin{center}
\includegraphics[scale=0.685]{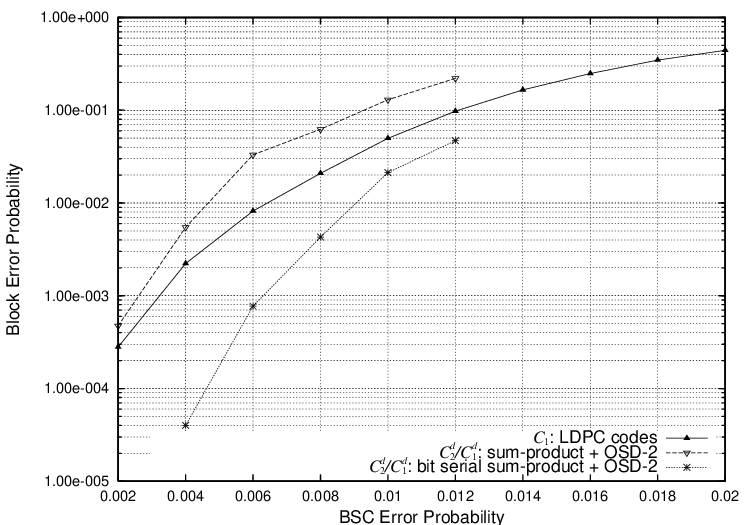}
 \caption{Decoding performance of the two codes constructed by (3,15) near regular LDPC codes of length 480 and rate 0.8; the rate as CSS codes is 0.6.}
\label{reg480}
\end{center}
\end{figure}

In the first example, we constructed the two codes by (3,15) near regular LDPC codes of length 480 and rate 0.8. 
The rate as CSS codes is 0.6. 
Fig. \ref{reg480} depicts the performance result for $C_1$ decoded by the general sum-product algorithm, 
$C_2^{\perp}(C_2^\perp /C_1^\perp)$ done by the original combined algorithm 
and by the modified one. In both cases of the original combined algorithm and the modified one, 
the algorithm of the removal of the cycles of length four is applied prior to sum-product decoding.  
We chose the maximum numbers of iterations of $C_1$ and $C_2^{\perp}(C_2^\perp /C_1^\perp)$ to be 100 and 256, respectively. 
The parameter of the OSD algorithm is set to order-2 reprocessing. 
In all Figs, henceforth, we note that $C^d$ represents a dual code $C^\perp$. 

In comparing the original combined algorithm and the modified one, 
a dramatic improvement with respect to the decoding performance is realized by our modified approach. 
The decoding performance of $C_2^\perp /C_1^\perp$ by the modified algorithm after the transformation of the Tanner graph
is superior to that of the LDPC code $C_1$ by sum-product decoding. 
  
\subsection{Approximative Decoding Approach}
The modified algorithm has a disadvantage that decoding complexity is too high to decode LDPC codes with block lengths of more than a few thousand bits. 
Therefore, we suggest and apply an approximative evaluation algorithm. 
For explaining the algorithm, we give one theorem: 

\begin{theorem}
\label{the}
On a binary erasure channel (BEC), sum-product decoding is equal to MLD after the transformation of Tanner graphs (or parity-check matrices) in response to erasure bits.
\end{theorem}
\begin{proof}
When the sum-product decoding algorithm fails on the BEC, a set of erasures is equal to the unique maximum stopping set \cite{DPRTU02}.
The stopping set $\mathcal{S}$ is the subset of the set of variable nodes in the Tanner graph, 
and all neighbors in $\mathcal{S}$ are connected to $\mathcal{S}$ at least twice. 
After the weights of the columns having an erasure bit are transformed to one, namely, the number of the connections is one, erasure bits can be decoded by the sum-product algorithm.  
If they cannot be transformed, the stopping set is a valid codeword. 
Erasure bits consisting of $\mathcal{S}$ containing valid codewords cannot be decoded even though we exploit MLD. 
Hence, sum-product decoding is equal to MLD after the transformation of the Tanner graphs in response to erasure bits.   
\end{proof}
High-density parity-check matrices have many small stopping sets due to many cycles, 
but transforming their Tanner graphs can remove small stopping sets as well as short cycles.  
\subsubsection{Decoding Process} 
\label{DP}
On the basis of Theorem \ref{the}, we evaluate the performance of $C_2^\perp(C_2^\perp /C_1^\perp)$ on the {\it BSC}, and the detailed procedure is performed as follows: 
\begin{enumerate}
\renewcommand{\labelenumi}{\alph{enumi})}
\item Transform the number of the edges of the nodes having an error in the Tanner graph to less than three. 
\item Remove cycles of length four in the Tanner graph. 
\item Decode received words to codewords by the general sum-product algorithm. 
\end{enumerate}
We anticipate some objections to this methodology, but in our computational simulations, 
decoding results have no failures when less than a few errors occur in the high-density part of parity-check matrices. 
To put it more precisely, received words are decoded to codewords in $C_2^\perp /C_1^\perp$, or they cannot be estimated to any codewords. 
In all cases that more than a few errors occur in the high-density part of parity-check matrices, no codewords can be estimated due to many cycles in Tanner graphs.   
For these reasons, only errors in the low-density part of parity-check matrices is important for the evaluation of $C_2^\perp(C_2^\perp /C_1^\perp)$,
and it is reasonable to suppose that this evaluation algorithm are near optimal on the BSC. 
\subsubsection{Generalized Decoding Process} 
If we generalize the methodology, then the decoding process is illustrated as follows: 
\begin{enumerate}
\renewcommand{\labelenumi}{\alph{enumi})}
\item Prepare a number of different parity-check matrices, composed of a low-density submatrix and a high-density submatrix, in the same code. 
\item Remove cycles of length four in each Tanner graph. 
\item Decode received words to codewords by sum-product algorithm using each parity-check matrix.  
\item Estimate codewords by a majority decoding result. 
\end{enumerate}
In practical, since we cannot prepare numerous parity-check matrices, 
we assume that the code $C_2^\perp$ is decoded by this generalized decoding process and, 
as mentioned in Sub-subsection \ref{DP}, evaluate the performance by transforming the Tanner graph in response to error bits with one matrix.
 
\begin{figure}[!t]
\centering
\includegraphics[scale=0.685]{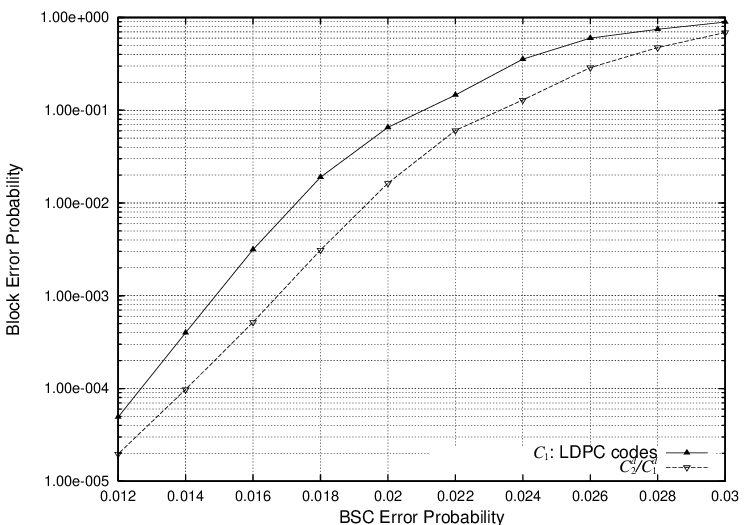}
\caption{Decoding performance of the two codes constructed by irregular LDPC codes of length almost 2000 and rate 0.78; the rate as CSS codes is 0.56. 
The maximum number of iterations was set to 100 and 256, respectively.}
\label{2000ir}
\end{figure}

\subsection{Decoding Performance of Moderate-Length Codes}
In this subsection, we show simulation results for moderate-length codes. 
As the way for the performance evaluation of $C_2^{\perp}(C_2^\perp /C_1^\perp)$, we applied the above approach. 

\begin{figure}[!t]
\begin{center}
\includegraphics[scale=0.685]{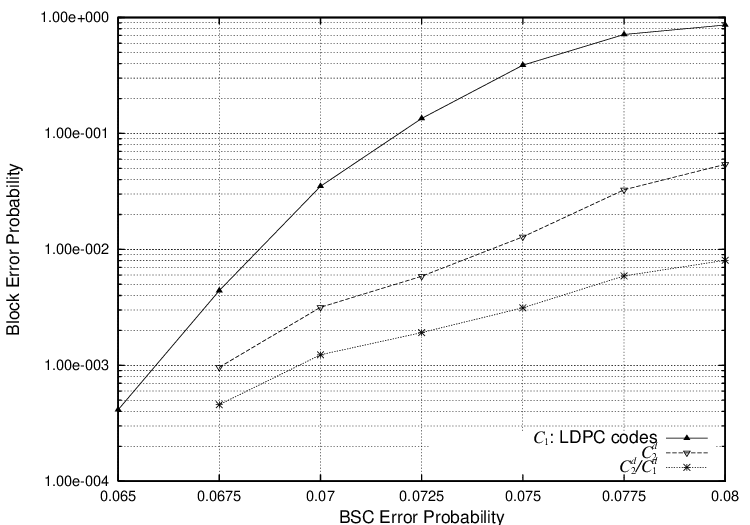}
 \caption{Decoding performance of the two codes constructed by irregular LDPC codes of length almost 8000 and rate 0.55; the rate as CSS codes is 0.1. 
The maximum number of iterations was set to 200 and 512, respectively.}
\label{8000ir}
\end{center}
\end{figure}

Figs. \ref{2000ir} and \ref{8000ir} depict the simulation results for the two codes by irregular LDPC codes in \cite{FOS05} of length 2183 and rate 0.78; length 7832 and rate 0.55, respectively. 
We employed 59 and 89 as the prime number $p$. 
Their LDPC codes were designed based on masking matrices given in App. \ref{AB}.
In both cases, the decoding performance of $C_2^{\perp}(C_2^\perp /C_1^\perp)$ is superior to that of the LDPC code $C_1$. 
Table \ref{coset} summarizes the relationship between the noise and the results covered in $C_2^\perp /C_1^\perp$ in Fig. \ref{8000ir}. 
We see from Table \ref{coset} that the coset $C_2^\perp /C_1^\perp$ covers the most codewords that are decoding failures in $C_2^\perp$. 
This is attributed to many small codewords in $C_1^\perp$, and it results from small row weights in $H_1$ of low-rate LDPC codes. 

\begin{table}[!t]
\begin{center}
\caption{Success rate covered in $C_2^\perp /C_1^\perp$ shown in Fig. \ref{8000ir}}
\label{coset}
\begin{tabular}{c|c}
\hline
Noise rate on BSC (\%)& Success rate covered in $C_2^\perp /C_1^\perp$ (\%)\\
\hline\hline
8.0 & 85.8\\
\hline
7.75 & 82.4\\
\hline
7.5 & 75.8\\
\hline
7.25 & 67.5\\
\hline
7.0 & 61.1\\
\hline
6.75 & 52.4\\
\hline
\end{tabular}
\end{center}
\end{table}

Since the code $C_2^{\perp}(C_2^\perp /C_1^\perp)$ has better performance than the LDPC code $C_1$ in every experimental result, 
the proposed method can bound Eve's mutual information on keys by the performance of the LDPC code $C_1$.    
As for Eve's mutual information on the generated 712 bit key in Fig. \ref{8000ir}, for example, they can be bounded by $I_{Eve}<0.5936$ with the crossover probability 6.5\% on the BSC 
and $I_{Eve}<6.312$ with  the crossover probability 6.75\%. 

If we choose LDPC codes of almost length 10000 and a little higher rate than 0.5 as the code $C_1$, 
then  it can be expected that the two codes are decodable with the crossover probability approximately 8.5\% on the BSC, keeping Eve's mutual information small.

\section{Conclusion}
\label{con}
In this paper, we introduced the simple method for constructing the good pair of codes using LDPC codes. 
First by the combined algorithm of robust decoding and the transformation of Tanner graphs, 
second by the approximative decoding approach for moderate-length codes, we evaluated the performance. 
Since the proposed method admits to apply arbitrary LDPC codes, 
it is less complex and better performance than existing methods. 
The simulation results indicate that the two codes constructed in this paper can be decodable against practical noise levels on the quantum channel 
and bound Eve's mutual information on keys by the performance of chosen LDPC codes. 
Therefore, this method allows secure and effective key generations.

\appendices
\section{}
\label{AA}
Masking matrices $W_{\rm rate}$ used in simulations are as follows:
{\tiny
\begin{equation*}
W_{0.82} =
\begin{bmatrix}
10000010000000001000100100001001000010001100110101101111\vspace{-.8mm}\\
10100000010000010001000000100011000100100011101011110110\vspace{-.8mm}\\
00000100100010000001001000100100010010001110110101011011\vspace{-.8mm}\\
01000100110000010000100000110001000100100011011010101001\vspace{-.8mm}\\
10010000100010010001000100010000000100010100101111010110\vspace{-.8mm}\\
01010010000010001000110001000100010001000111010000111011\vspace{-.8mm}\\
00000101010100100010000101001000001000001001010111010101\vspace{-.8mm}\\
00100001000101001100010010010000010001100110001010101110\vspace{-.8mm}\\
00001000001001100000001001000010101000010011100111101101\vspace{-.8mm}\\
01000010001001000110010000001000100010000100111001111010\vspace{-.8mm}\\
00011000000100000010000010000110001000010011111110010111\vspace{-.8mm}\\
00101001001000100100001010000000100001000101001111111101\vspace{-.5mm}
\end{bmatrix}{\normalsize ,}
\end{equation*}
\begin{equation*}
W_{3/4} =
\begin{bmatrix}
000001000000100001000001000010101011010110101011011\vspace{-.8mm}\\
000000000100001001010000010000110101101011010101101\vspace{-.8mm}\\
100000100001000000100100000001101011010011101010110\vspace{-.8mm}\\
100001000010000100000100000000101101010110011101011\vspace{-.8mm}\\
000000100000100000010000100101011010110101101010110\vspace{-.8mm}\\
000010000010000010000001010000101101011100101101011\vspace{-.8mm}\\
010100010001000000000000100001011010110101011010110\vspace{-.8mm}\\
001000001000010000010000001011010110101101010110010\vspace{-.8mm}\\
000100000100000010000100001000110101101010110101101\vspace{-.8mm}\\
100000100000010001000010000001101010110101101011010\vspace{-.8mm}\\
000010000100000100000010000010110101011010110101101\vspace{-.8mm}\\
010000010000010000100001000001011010101101011010110\vspace{-.8mm}\\
000010000010001000001000001000101101011010110101101\vspace{-.8mm}\\
001000001000001000100000100001010110101011010110101\vspace{-.8mm}\\
000100001000000100001000000101010101101011100111010\vspace{-.8mm}\\
001000010000100000001000010001010110101101011010101\vspace{-.8mm}\\
010001000001000010000010000101101011010110101101001\vspace{-0.5mm}
\end{bmatrix}{\normalsize ,}
\end{equation*}
\begin{equation*}
W_{2/3} =
\begin{bmatrix}
0000000100000010100000000000010101001010100101\vspace{-0.8mm}\\
0000100001000101000000000000101001010101001010\vspace{-0.8mm}\\
0000000010000000100000000101010010101010010101\vspace{-0.8mm}\\
0100000000000010000001000010101010010101001010\vspace{-0.8mm}\\
0000010000100100000000000101010001101001010100\vspace{-0.8mm}\\
0000000100000000010000000011010100101010100101\vspace{-0.8mm}\\
0010000001000000001000000100101010010101001010\vspace{-0.8mm}\\
1000000010000000010000000001010100101010010101\vspace{-0.8mm}\\
0001000000100000000100001000101001010100101010\vspace{-0.8mm}\\
0000000000010000001000100011001010100101010100\vspace{-0.8mm}\\
0000000001000000000100011000010100101001010101\vspace{-0.8mm}\\
1001001000001000000000000000101001010100101010\vspace{-0.8mm}\\
0100000000010000000000000001110010100101010010\vspace{-0.8mm}\\
0000000000001000001000001000101010010011010001\vspace{-0.8mm}\\
0000100000010000000010000001000101001100101010\vspace{-0.8mm}\\
0000010000001000000100000000101001010101010001\vspace{-0.8mm}\\
0010001000000000010010000001000110100010101010\vspace{-0.8mm}\\
0000010000000001000001010000100101010010101001\vspace{-0.8mm}\\
0100000100000100000000100001001010101001010100\vspace{-0.8mm}\\
0010101000000000000000010000010101010010100110\vspace{-0.8mm}\\
0001000000100000100000100000010110001100101001\vspace{-0.8mm}\\
1000000000000010000001000000101011010010101010\vspace{-0.8mm}\\
0000000010000001000010000001010100101010010101\vspace{-0.5mm}
\end{bmatrix}{\normalsize ,}
\end{equation*}
\begin{equation*}
W_{0.55} =
\begin{bmatrix}
00000100001000000000000010000101000001\vspace{-.8mm}\\
00001000000100000010100000100010000000\vspace{-.8mm}\\
00000000100000000101000000001000101000\vspace{-.8mm}\\
00001000000010001000000100000000010010\vspace{-.8mm}\\
10000000010000000000011000010000000100\vspace{-.8mm}\\
00100010000001000000000000001000001010\vspace{-.8mm}\\
00100100000010000000100000100010000000\vspace{-.8mm}\\
00010000000101000000000000000101001000\vspace{-.8mm}\\
00000010000000001001010000000000010010\vspace{-.8mm}\\
10001000010000000010000001000100000000\vspace{-.8mm}\\
01000000000100000000100100010001000000\vspace{-.8mm}\\
00000000101000010000000010000010000001\vspace{-.8mm}\\
00100000000010001000001000000000101000\vspace{-.8mm}\\
00001000000001010000000000010100010000\vspace{-.8mm}\\
00010000000100000000010000101000000001\vspace{-.8mm}\\
00000001100000100100000100000001000000\vspace{-.8mm}\\
10000010001000000001001000000010000000\vspace{-.8mm}\\
01000100010000000010000000100000000100\vspace{-.8mm}\\
01000000000000100100000010010010000000\vspace{-.8mm}\\
10000000100000010000100000000100100000\vspace{-.8mm}\\
00000100000001001001000001000000010000\vspace{-.8mm}\\
00000011010010000000000010000000100000\vspace{-.8mm}\\
00010000001000010010010000001000000010\vspace{-.8mm}\\
00100001000000000000001001000001001000\vspace{-.8mm}\\
01000000001000000000010100000000100100\vspace{-.8mm}\\
00010000010000001000001000100000010000\vspace{-.8mm}\\
00100000100100100000100001000000000100\vspace{-.8mm}\\
00010001000000100101000000001000000001\vspace{-.8mm}\\
00001000000010010000000100010000000010\vspace{-.8mm}\\
10000010000001000100000010000000000100\vspace{-.8mm}\\
01000101000000100010000001000000000001\vspace{-.5mm}
\end{bmatrix} {\normalsize .}
\end{equation*}
}
\section{}
\label{AB}
Masking matrices $W_{\rm rate}$ used in simulations are as follows:
{\footnotesize
\begin{eqnarray*}
W_{0.8} =
\begin{bmatrix}
00010101011000111010110110101 \vspace{-1.0mm}\\
00100100110111111011010010110 \vspace{-1.0mm}\\
10001010010100110111011011110 \vspace{-1.0mm}\\
01100010011010011101101101001 \vspace{-1.0mm}\\
10011001001111101000101011110 \vspace{-1.0mm}\\
01001010001101010111110101101 \vspace{-1.0mm}\\
01010001110111000111011100011 \vspace{-1.0mm}\\
10100100101011101100101111011 \vspace{-.5mm}
\end{bmatrix}{\normalsize ,}
\end{eqnarray*}
}
{\tiny
\begin{eqnarray*}
W_{0.55} =
\begin{bmatrix}
001000001000000000000000010000000000100000000010 \vspace{-0.8mm}\\
000100000000000000000000000000100010010001000001 \vspace{-0.8mm}\\
010010000000000001001100000000000001000000000000 \vspace{-0.8mm}\\
100000100100000000000001000001000000000000000100 \vspace{-0.8mm}\\
000000000010010000010000000010000000000000110000 \vspace{-0.8mm}\\
000000010000100010100010000000010000000000000000 \vspace{-0.8mm}\\
000000000000001100000000001000001100000010000000 \vspace{-0.8mm}\\
000001000001000000000000100100000000001000001000 \vspace{-0.8mm}\\
000100100000000000000001000000101001000000000000 \vspace{-0.8mm}\\
000000000010000001000000000000010000011010000000 \vspace{-0.8mm}\\
000010000000001000100010010000000000000000100000 \vspace{-0.8mm}\\
000000010000010000000100001100000000000100000000 \vspace{-0.8mm}\\
000000000001000010000000000000000110000000010100 \vspace{-0.8mm}\\
000000000000100000010000100000000000100001000001 \vspace{-0.8mm}\\
101000000000000100001000000001000000000000001000 \vspace{-0.8mm}\\
010001001100000000000000000010000000000000000010 \vspace{-0.8mm}\\
000010000000000010000000010001000010000000000010 \vspace{-0.8mm}\\
000000000000000100001000000000000000011010000001 \vspace{-0.8mm}\\
101100010000010001000000000000000000000000000000 \vspace{-0.8mm}\\
000000100000000000000000000010001100000001010000 \vspace{-0.8mm}\\
000001000010100000010000100000100000000000000000 \vspace{-0.8mm}\\
010000000101000000000001000000010000000000100000 \vspace{-0.8mm}\\
000000000000001000100010000100000000100000000100 \vspace{-0.8mm}\\
000000001000000000000100001000000001000100001000 \vspace{-0.8mm}\\
000100100000000000001000010001000100000000000000 \vspace{-0.8mm}\\
010000000000010000000000100000000010001000000010 \vspace{-0.8mm}\\
001001000011000000000001000100000000000000000000 \vspace{-0.8mm}\\
000000000000001000000100000000111000000000001000 \vspace{-0.8mm}\\
000010000000000010110000000010000000000010000000 \vspace{-0.8mm}\\
000000001000100000000000001000000000000001100001 \vspace{-0.8mm}\\
000000000100000001000010000000000000110100000000 \vspace{-0.8mm}\\
100000010000000100000000000000000001000000010100 \vspace{-0.8mm}\\
000000010001000100000000000001000001000000100000 \vspace{-0.8mm}\\
010000101000000000000000000000100100000000001000 \vspace{-0.8mm}\\
000010000000000010000000001000001000001010000000 \vspace{-0.8mm}\\
001000000000000000100010000000000000100000010010 \vspace{-0.8mm}\\
100000000100000000001001010000010000000000000000 \vspace{-0.8mm}\\
000000000000111000000000000000000000010001000001 \vspace{-0.8mm}\\
000100000010000000010100000000000010000100000000 \vspace{-0.8mm}\\
000001000000000001000000100110000000000100000100 \vspace{-0.5mm}
\end{bmatrix}{\normalsize .}
\end{eqnarray*}
}Almost all matrices in App. \ref{AA} and \ref{AB} were designed based on degree distribution pairs given in \cite{AU}.

%\section*{Acknowledgment}

%\balance

\balance
\end{document}